%% file: inertia4arXiv.tex
\documentclass[letterpaper,10pt,twocolumn,final,journal,oneside]{IEEEtran}

\usepackage{hyperref}
\usepackage{amsmath}
\usepackage{amsfonts}
\usepackage{dsfont}
\usepackage{graphics}
\usepackage{psfrag}
\usepackage{array}
\usepackage{shortvrb}
\usepackage{epsf}
\usepackage{nicefrac}
\usepackage{rotating}
\usepackage{empheq}
\usepackage{multicol}
\usepackage{acronym}
\usepackage{enumerate}
\usepackage[Gray,amssymb]{SIunits}
\usepackage{booktabs}
\usepackage{multirow}
\usepackage{amssymb}
\usepackage{pifont}
\usepackage{multicol}
\usepackage{color}
\usepackage[normalem]{ulem}
\usepackage[utf8]{inputenc}
\usepackage[T1]{fontenc}
\usepackage{tikz}
\usepackage{stfloats}
\usepackage{ccicons}
\usepackage{balance}
\newcommand{\circled}[1]{\raisebox{.5pt}{\textcircled{\raisebox{-.9pt} {#1}}}}	

\newcommand{\pu}{\text{[pu]}}
\definecolor{brown}{rgb}{0.59, 0.29, 0.0}
\definecolor{amethyst}{rgb}{0.6, 0.4, 0.8}
\definecolor{emerald}{rgb}{0.31, 0.78, 0.47}
\definecolor{gamboge}{rgb}{0.89, 0.61, 0.06}
\definecolor{green(ryb)}{rgb}{0.4, 0.69, 0.2}
\definecolor{icterine}{rgb}{0.99, 0.97, 0.37}
\definecolor{amber}{rgb}{1.0, 0.75, 0.0}
\definecolor{darkolivegreen}{rgb}{0.33, 0.42, 0.18}
\definecolor{magenta(process)}{rgb}{1.0, 0.0, 0.56}

\newcommand{\red}[1]{\textcolor{red}{#1}}

\newcommand{\bb}{$\bullet$}

\newcommand\redsout{\bgroup\markoverwith{\textcolor{red}{\rule[0.5ex]{2pt}{0.4pt}}}\ULon}

\newcommand{\RefFig}[1]{Fig.~\ref{#1}}

\newcommand{\InRefEq}[1]{Equation~(\ref{#1})}

\newcommand{\RefEq}[1]{Eq.~(\ref{#1})}

\newcommand{\Refe}[1]{(\ref{#1})}

\acrodef{tg}[\textsc{tg}]{turbine governor}
\acrodef{avr}[\textsc{avr}]{automatic voltage regulator}
\acrodef{coi}[\textsc{coi}]{center of inertia}
\acrodef{res}[\textsc{res}]{renewable energy sources}
\acrodef{pv}[\textsc{pv}]{photovoltaic}
\acrodef{mpp}[\textsc{mpp}]{maximum power point}
\acrodef{agc}[\textsc{agc}]{automatic generation control}
\acrodef{pf}[\textsc{pf}]{power flow}
\acrodef{gbts}[\textsc{gbts}]{Great Britain Transmission System}

\IEEEoverridecommandlockouts

\newcommand\copyrighttext{%
  \footnotesize
  \centering\ccbyncnd\\ \copyright~2021. This manuscript version is made available under the\\
  Creative Commons Attribution-NonCommercial-NoDerivatives 4.0 International License
  \href{http://creativecommons.org/licenses/by-nc-nd/4.0/}{http://creativecommons.org/licenses/by-nc-nd/4.0/.}\\
  DOI: \href{https://doi.org/10.1016/j.ijepes.2021.106842}{10.1016/j.ijepes.2021.106842}}
\newcommand\copyrightnotice{%
\begin{tikzpicture}[remember picture,overlay]
\node[anchor=south,yshift=0pt] at (current page.south) {\setlength{\fboxrule}{0pt}\fbox{\parbox{\dimexpr\textwidth-\fboxsep-\fboxrule\relax}{\copyrighttext}}};
\end{tikzpicture}%
}

\begin{document}

\title{Effects of Inertia, Load Damping and Dead-Bands on
Frequency Histograms and Frequency Control of Power Systems}

\author{Davide~del~Giudice,~\IEEEmembership{Member,~IEEE,} Angelo~Brambilla,~\IEEEmembership{Member,~IEEE,} Samuele~Grillo,~\IEEEmembership{Senior~Member,~IEEE,} and Federico~Bizzarri,~\IEEEmembership{Senior~Member,~IEEE}%
\thanks{D. del Giudice, A. Brambilla, S. Grillo, and F. Bizzarri are Dipartimento di Elettronica, Informazione e Bioingegneria, Politecnico di Milano, p.za Leonardo da Vinci, 32, I20133, Milano, Italy.}
\thanks{F. Bizzarri is also with Advanced Research Center on Electronic Systems ``E. De Castro'' (ARCES), University of Bologna, Italy.}
\thanks{Corresponding author: Angelo Brambilla (angelo.brambilla@polimi.it).}}

\IEEEaftertitletext{\copyrightnotice\vspace{1.1\baselineskip}}
\maketitle

\begin{abstract}
The increasing penetration of renewable energy sources has been leading to the progressive phase-out of synchronous generators, which constitute the main source of frequency stability for electric power systems. In the light of these changes, over the past years, some power systems started to exhibit an odd frequency distribution characterised by a bimodal behavior. This results in increased wear and tear of turbine governors and, in general, in degraded frequency performances. This is a cause of concern for grid operators, which have become increasingly interested in understanding the factors shaping frequency distribution. This paper explores the root causes of unwanted frequency distributions. The influence of some main aggregate system parameters on frequency distribution is detailed. The paper also shows that the implementation of the so-called synthetic inertia can lead to a robust unimodal frequency distribution.
\end{abstract}


\section{Introduction}
\IEEEPARstart{In}{ electric} power systems, frequency is the most important parameter
that transmission system operators (\textsc{tso}s) must control and
manage, trying to limit its variations within strict bounds \cite{Kundur:1994}.
Several corporations, as the North American Electric
Reliability Corporation (NERC) \cite{NERC:2011}, prepared documents concerning
the decreasing trend in the frequency quality of a complex power system \cite{NERC:2012}.
To solve this issue, some counteractions and remedies have been proposed,
e.g., new settings for traditional generators dead-bands and droop gains \cite{NERC:2011},
the exploitation of \ac{res} \cite{Kroposki:2017} and storage systems
for frequency regulation services \cite{ARRIGO2020105428},
demand response, the introduction
of load and generation aggregators, and the adoption of thermostatically controlled loads,
which can implement controllers that link the absorbed power to
frequency variations and can mimic an inertial response.
\cite{Obaid:2019}.
There is quite a large consensus that the main reason for this
trend is the continuous reduction of system inertia, which is primarily
due to the increasing penetration of \ac{res} and converter-interfaced
loads \cite{Ulbig:2014}. The focus is on these components since they deliver/absorb power
regardless of frequency variations, thereby constituting frequency-independent
generators/loads.
The general indication is that the decrease in system inertia has to be restored
in some way as the number of conventional synchronous generators
are replaced by \ac{res} \cite{Tielens:2016}.
For instance, this could be done by implementing the so-called synthetic
(or virtual) inertia in converter-interfaced equipment, which allows
mimicking the frequency response of synchronous generators following
power imbalances \cite{Tamrakar:2017}.
While on the one hand, the factors determining the frequency behavior
following a power disturbance are already well established, on the
other hand the analysis of the parameters shaping power system
frequency distribution due to the presence of stochastic load
variations is still at an early stage of research. Early papers and reports
addressing this topic showed that frequency distributes around its nominal
value in a way that leads to {\em bimodal} histograms, which are
characterised by two frequency occurrence peaks, often located almost
symmetrically around the nominal frequency value.
Such an odd frequency behavior has been reported in North America
\cite{NERC:2019}, Ireland \cite{7540970}, and, lastly, Great Britain \cite{8626538},
 whose frequency measurement data are publicly available in \cite{GB:2016}.
This bimodal behavior is in contrast to the assumption typically
put forward for the design of system controllers and statistical analysis,
according to which frequency distribution is unimodal. Moreover, as shown in the
following, it can cause an increased tear and wear of turbine governors. As a
consequence, such a distribution is undesirable.

\subsection{Contributions}
Starting from these considerations, we investigate and identify
the reasons that lead to unwanted frequency
distributions such as bimodal frequency histograms.
This is initially done analytically through a formal approach applied
to a simplified, {\em linear} power system model comprising
a stochastic load \cite{6547228},
which constitutes the source of power mismatches.
We then use a modified version of the well-known \textsc{ieee 14 bus} power
system as a benchmark.
Detailed numerical simulations are performed to compute frequency
distributions in different operating scenarios.
When possible, results are compared with experimental measurements of
more complex power systems.

Some considerations and guidelines on how to act on power systems
to avoid a bimodal frequency distribution are proposed.
These considerations take into account effects due to
dead-bands of \acp{tg}, \ac{agc}, the inertia of synchronous generators,
variability, and stochastic characteristics of loads and generation,
elements with power versus frequency dependence
(e.g., frequency-dependent loads) and possibly synthetic inertia.

The main results we show are: (i) simply increasing or restoring
inertia without considering the aggregate value of the load damping
parameter can be ineffective.
This is in contrast to the common belief that an increase in inertia is
beneficial per se.
(ii) Bimodal frequency distribution histograms are due to at least two
remarkably different reasons.
The former is an inappropriate procedure to aggregate frequency samples
that simply leads to an ``artifact''
\cite{8626538}.
The latter is due to the adoption of ineffective dead-bands, as well as of an
unsuitable choice of the values of other system parameters, which lead to
continuous wear and tear of \acp{tg} and poorly-controlled swings
following long-term, small power imbalances.
(iii) The introduction of adequate controllers in electronically-interfaced
loads and \ac{res} generation can sensibly and beneficially modify the
aggregate values of system parameters and mainly load damping and inertia.
We show that this leads to very narrow, unimodal frequency distributions.

\section{Simplified model of a power system with stochastic load}
\label{S:LoadModel}
Let us assume to have a simplified and linear power system model.
In such a system, we consider $\Delta \omega$ as the deviation from
$1\,\pu$ of the angular frequency of the \ac{coi}.
We model it through the swing equation
\begin{equation}
2H \frac{d\Delta\omega}{dt} = - D_{\mathrm{L}}  \Delta\omega -
\eta(t) ~,
\label{E:SimpleSwing}
\end{equation}
where $H$ is the inertia of the \ac{coi} and $D_{\mathrm{L}}$
is the load damping coefficient.
We assume that the $\eta(t)$ power fluctuations in the swing equation are
due to a stochastic load, modeled by the linear stochastic differential
equation
\begin{equation}
d\eta = \left( \mu - \alpha \eta(t) \right)dt + b\,dW_t ~,
\label{E:Oup}
\end{equation}
where $\alpha$ is known as the reciprocal of the load reversal time, $b$
governs the variance of $\eta(t)$ and $W_t$ is
a scalar Wiener process.
The use of a stochastic model of power loads is not novel;
see for example \cite{5298967,574922,DELGADO2014267}.

The choice of a linear model is plausible when the dead-bands of
\acp{tg} are wide enough with respect to the frequencies deviations
caused by the stochastic load. If so, the intervention of \acp{tg} is prevented.
Furthermore, when the stochastic load leads to modest frequency variations,
a linearised version of a complex non-linear power system can be employed.
The linear model in \eqref{E:SimpleSwing} is based on the \ac{coi} and
considers an aggregate, unique swing equation \footnote{
The assumption that a single \ac{coi}
is a good simplified representative of a complex power system holds in a
large number of cases but not always.
Frequency may vary substantially between locations in some complex,
extended grids with possibly weekly coupled areas due to transient
and oscillatory dynamics in the network.
Our single \ac{coi} simplification allows us to derive analytical
expressions that give insight on how aggregate systems parameters act
on frequency deviations.
These analytical expressions significantly grow in complexity if more areas
are considered.} \cite{8626538,Kundur:1994}.

Equations \Refe{E:SimpleSwing} and \Refe{E:Oup} can be written in compact
form as
\begin{equation}
\left[\begin{IEEEeqnarraybox}[][c]{,c,}
d\Delta\omega\\
d\eta%
\end{IEEEeqnarraybox}\right]
=
\left[\begin{IEEEeqnarraybox}[][c]{,c,}
0\\
\mu%
\end{IEEEeqnarraybox}\right]
dt -
\left[\begin{IEEEeqnarraybox}[][c]{,c/c,}
\frac{D_{\mathrm{L}}}{2H} & \frac{1}{2H} \\
0                         & \alpha %
\end{IEEEeqnarraybox}\right]
\left[\begin{IEEEeqnarraybox}[][c]{,c,}
\Delta\omega\\
\eta%
\end{IEEEeqnarraybox}\right]
dt +
\left[\begin{IEEEeqnarraybox}[][c]{,c,}
0\\
b%
\end{IEEEeqnarraybox}\right]
d {W}_t
\label{E:SimSys}~.
\end{equation}
\RefEq{E:SimSys} belongs to the general class of multi-dimensional
linear stochastic equations in the form \cite{Arnold1974}
\begin{equation}
\begin{aligned}
d \mathbf{x} &= \left( \boldsymbol{\mu}(t) -\mathbf{A}\right) \mathbf{x} \,dt +
\mathbf{B}\, d\mathbf{W}_t\\
\mathbf{x}(t_0) &= \mathbf{x}_o~,
\end{aligned}
\label{E:Ou}
\end{equation}
where $\mathbf{W}_t$ is an $M$-dimensional Wiener process,
$\mathbf{A} \in \mathds{R}^{N \times N}$,
$\mathbf{B} \in \mathds{R}^{N \times M}$ are constant
matrices, $\boldsymbol{\mu}(t)$ is
an $N$-dimensional vector of time functions, and $\mathbf{x}_o$
is the initial condition. \InRefEq{E:Ou} admits the solution
\begin{equation}
\mathbf{x}(t) = e^{-\mathbf{A}(t-t_0)} \mathbf{x}_o + \!\!
    \int_{t_0}^{t}{\!\!e^{-\mathbf{A}(t_0-s)}\!\left( \boldsymbol{\mu}(s) ds +\!
        \mathbf{B} d\mathbf{W}_s \right)}
~,
\label{E:OuMean}
\end{equation}
and
\begin{equation}
\mathrm{E}\left[\mathbf{x}(t)\right] = e^{-\mathbf{A}(t-t_0)}
  \left( \mathrm{E}[\mathbf{x}_o] +
    \int_{t_0}^{t}{\!\!e^{-\mathbf{A}(t_0-s)} \boldsymbol{\mu}(s) ds} \right)
\label{E:Ave1}~.
\end{equation}
Without loss of generality, in the following we assume $t_0=0$ and $\mathrm{E}[\mathbf{x}_o] = 0$.

The covariance matrix of $\mathbf{x}(t)$ is the solution of the
\begin{equation}
\dot{\mathbf{w}} =
- \left( \mathbf{A} \mathbf{w} + \mathbf{w} \mathbf{A}^{\mathrm{T}} \right) +
\mathbf{B} \mathbf{B}^{\mathrm{T}}
\label{E:OuCov}
\end{equation}
differential equation (the $\mbox{}^\mathrm{T}$ super-script denotes
transposition).
In our analysis, we focused on \Refe{E:SimSys} letting both
${\mu}$ constant and ${\mu(t)=\rho \sin(\Psi t)}$.
In the first case, \Refe{E:Oup} models the evolution of the
Ornstein-Uhlenbeck process \cite{PhysRevE.54.2084}, which is
exponentially-autocorrelated and normally-distributed,
with $\displaystyle{\mathrm{E}[\eta(t)] =
\nicefrac{\mu}{\alpha} + \left( \eta(0) - \nicefrac{\mu}{\alpha} \right)
e^{-\alpha t}}$
time-varying mean and
\begin{equation}
\mathrm{Var}[\eta(t)] = \frac{b^2}{2\alpha}\left( 1 - e^{-2\alpha t}\right)
\label{E:OuVar}
\end{equation}
time-varying variance.
$\eta(t)$ asymptotically tends to
${\nicefrac{\mu}{\alpha}}$ as
$t$ tends to infinity.
Since, in particular, we assume ${\mu}=0$,
as far as $\alpha t  \gg 0$ the stochastic load on average does not
contribute power and ${\mathrm{E}}[\Delta \omega]_{\alpha t  \gg 0}$ is null too.

In the second case, $\mu(t)$ is used to model
a slowly-varying deterministic and periodic (daily) power imbalance.
We assume ${\Psi \ll 2\pi \alpha}$, viz. the period
of $\mu(t)$ is much larger than the process reversal-time.
By exploiting \Refe{E:Ave1} with $2H\alpha \neq D_{\mathrm{L}}$ we get
\begin{equation}
\begin{array}{lcl}
{\mathrm{E}}[\Delta \omega] &\!\!=\!\!&
\displaystyle{
k e^{-\frac{D_{\mathrm{L}}}{H}t} +
\frac{\rho \Psi e^{-\alpha t} }{\left(\alpha^2 + \Psi^2\right)
  \left(2H\alpha - D_{\mathrm{L}}\right)}~+} \\[3mm]
&\!\!+\!\!& \displaystyle{\rho_c \cos\left( \Psi t \right) +
  \rho_s \sin\left( \Psi t \right)}
\end{array}
\label{E:Mean1}~,
\end{equation}
where
\begin{equation*}
\begin{array}{l}
\displaystyle{\rho_c = \frac{\rho \Psi
\left(2H\alpha + D_{\mathrm{L}}\right)}
{\left(4H^2 \Psi^2 + D_{\mathrm{L}}^2\right) \left( \alpha^2 + \Psi^2\right)}}\\[3mm]
\displaystyle{\rho_s =
  \frac{\rho\left(2H \Psi^2 - \alpha D_{\mathrm{L}}\right)}
  {\left(\alpha^2 + \Psi^2 \right)\left(4H^2\Psi^2 + D_{\mathrm{L}}^2\right)}}
\end{array}
\end{equation*}
and $k$ is chosen to guarantee that $\mathrm{E}[\Delta \omega(0)] = 0$.
By solving \Refe{E:OuCov} the expression of the variance of $\Delta \omega$ is
\begin{equation}
\mathrm{Var}[\Delta \omega] \!=\!
\sigma^2\!\left( 1\!-\!\kappa_1 e^{-\frac{D_{\mathrm{L}}}{H}t}\!+\!
\kappa_2 e^{-\frac{D_{\mathrm{L}}+2H\alpha}{2H}t}\right)\!-\!
\kappa_3 e^{-2\alpha t}
~,
\label{E:SigmaOmega1}
\end{equation}
where
\begin{equation}
\sigma =
\frac{b}{\sqrt{2\alpha D_{\mathrm{L}}\left( D_{\mathrm{L}} + 2H\alpha\right)}}
\label{E:Std0}
\end{equation}
and
\begin{equation*}
\begin{array}{l}
\displaystyle{\kappa_1 =
  \frac{2 H \alpha\left( D_{\mathrm{L}} + 2 H \alpha\right)}
  {\left(D_{\mathrm{L}} -2 H \alpha\right)^2}} \\[4mm]
\displaystyle{\kappa_2 =
  \frac{8H \alpha D_{\mathrm{L}}}{\left(D_{\mathrm{L}} -2 H \alpha\right)^2}} \\[3mm]
\displaystyle{\kappa_3 =
  \frac{b^2}{2\alpha\left(D_{\mathrm{L}} -2 H \alpha\right)^2}}
~.
\end{array}
\end{equation*}

An important aspect is that $\boldsymbol{\mu}(t)$ does not appear in
\Refe{E:OuCov}, thus $\mathbf{w}(t)$ does not depend on it. Consequently,
\Refe{E:Std0} is invariant w.r.t. ${\mu}(t)$.
When $t$ is sufficiently large, \Refe{E:Mean1} and \Refe{E:SigmaOmega1}
simplify as
\begin{equation}
\mathrm{E}[\Delta \omega]_{\alpha t \gg 0} \simeq
  \rho_c \cos\left( \Psi t \right) +
  \rho_s \sin\left( \Psi t \right)
\label{E:Mean2}~,
\end{equation}
viz. a periodic function with period equal to $\nicefrac{2\pi}{\Psi}$, and
\begin{equation}
\mathrm{Var}[\Delta \omega]_{\alpha t \gg 0} \simeq \sigma^2~.
\label{E:Std}
\end{equation}
By observing \Refe{E:Std0}, we see that $\sigma$ is a constant which
is inversely proportional to the ${D_{\mathrm{L}}}$ load damping, to
the $\nicefrac{1}{\alpha}$ load reversal time, and the $H$ inertia.
It is worth pointing out that the $H$ inertia is multiplied by
$\alpha$ at the denominator of $\sigma$ and then by $D_{\mathrm{L}}$.
This means that {the smaller $D_{\mathrm{L}}$, the lower the effectiveness
of inertia in limiting frequency deviations}.
This is in contrast to the common idea that an increment of the overall
system inertia is beneficial, regardless of the value of other aggregate
system parameters.
From our standpoint, this is a relevant result.
We will consider this matter in further detail in the following.

Another aspect is that {$D_{\mathrm{L}}$ adds to $2H\alpha$ and it is thus
deemed to play a significant role in limiting frequency variations}.
In the light of this, $D_{\mathrm{L}}$ could be increased by exploiting
thermostatically controlled loads that link frequency variations to the
absorbed power \cite{7731152,4275780,6832599,8586247,7579133},
as anticipated in the Introduction, as well as
load damping and synthetic inertia, which require a suitable implementation
of \ac{res} generation and storage systems
\cite{jibji2019frequency,fang2018improved,kim2019supercapacitor}.

We pay now some more attention to $\sigma$. As $D_{\mathrm{L}}$ tends to $0$, $\sigma$ increases, i.e., the distribution of $\Delta \omega$ around $0$ flattens and extends.
Thus, as shown in the following, the reduction of $D_{\mathrm{L}}$ leads to an increased probability of crossing the dead-bands implemented in \acp{tg}.
The variance of $\Delta \omega$, in the extreme case when $D_{\mathrm{L}}
 =0$, is
\begin{equation}
\mathrm{Var}[\Delta \omega]_{D_{\mathrm{L}}
 =0} = \frac{b^2}{4H^2 \alpha^2}\!
  \left[t\!+\!\frac{1}{\alpha} \!\!
  \left(2e^{-\alpha t}\!-\frac{1}{2}e^{-2\alpha t}\!-\frac{3}{2}\right)\!
  \right]
~.
\label{E:Dl0H}
\end{equation}
This expression clearly shows that $\mathrm{Var}[\Delta \omega]_{D_{\mathrm{L}}
 =0}$ increases with time.
As a consequence, the probability that dead-bands are crossed increases with time too.
When $t$ is sufficiently large so that $\alpha t$ is far larger than 0,
the two exponential terms in \Refe{E:Dl0H} are practically 0
and
\begin{equation}
\displaystyle{\mathrm{Var}[\Delta \omega]_{D_{\mathrm{L}}
 =0, \alpha t \gg 0} \, \simeq \, \frac{b^2}{4H^2\alpha^2} \left(t-\frac{3}{2\alpha}\right)}
~.
\label{E:Dl0H2}
\end{equation}
It means that, at least in the simplified model of the power system,
if $D_{\mathrm{L}}=0$, by acting on the system inertia alone,
the frequency is likely to trespass the bounds of the dead-bands of
the \acp{tg}, anyway.
It is the first relevant result of the paper.
We dare say that this is in contrast to the common belief that a
sufficiently high increase of the aggregate system inertia (\ac{coi} inertia)
is always beneficial to lower frequency variations.
It seems false with stochastic loads.
We come back on this aspect in Section \ref{S:Inertia}.

\section{Numerical simulation setup}
\label{S:NumSetUp}
As previously said, the simplified model described by
\Refe{E:SimSys} may grab the main features of a complex non-linear
power system model that works in suitable conditions.
To show what happens to frequency when a complex power system
is stimulated by loads modeled as stochastic and how frequency depends
on system settings and aggregated parameters,
we resort to numerical simulations and elect the well known
\textsc{ieee 14 bus} power system to the role of our benchmark\footnote{
Simulations were performed with the simulator \textsc{pan}
\cite{8052254}.}.
Since the insight coming from \Refe{E:SimSys} has general validity,
we could have chosen other known power system models than the
\textsc{ieee 14 bus} power system.
%%%%%%%%%%%%%%% BEGIN FIGURE %%%%%%%%%%%%%%%
\begin{figure}[h]
\centering
\includegraphics[width=\columnwidth]{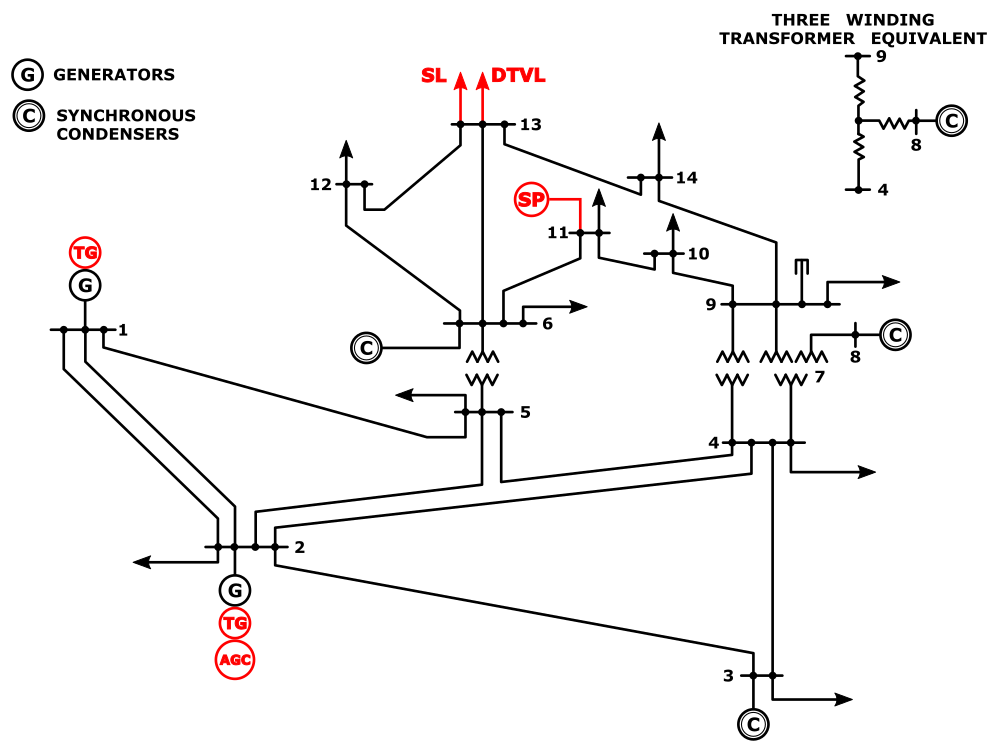}
\caption{One-line diagram of the modified \textsc{ieee} 14-bus system. \textsc{g}: generator.
\textsc{c}: synchronous condenser. \textsc{tg}: turbine governor. \textsc{agc}: automatic generation control.
\textsc{sl}: stochastic load. \textsc{dtvl}: daily time-varying load. \textsc{sp}: solar plant. The parameter values of the standard \textsc{ieee} 14-bus system can be found in \cite{Milano:2010} (see Chapter 21.4).
\label{F:14bus}}
\end{figure}
%%%%%%%%%%%%%%% END FIGURE %%%%%%%%%%%%%%%

In the following, the \textsc{ieee 14 bus} power system is extended with
several additions (see \RefFig{F:14bus}), which
are used in different ways and scenarios depending on the factors shaping
frequency deviations that we want to highlight.
\begin{itemize}
\item
Each of the two generating units of the \textsc{ieee 14 bus} power system is
equipped with a \ac{tg}, whose control scheme is modified by including
a dead-band block (see \RefFig{F:TgI}).
%%%%%%%%%%%%%%% BEGIN FIGURE %%%%%%%%%%%%%%%
\begin{figure}[h]
\centering
\includegraphics[width=\columnwidth]{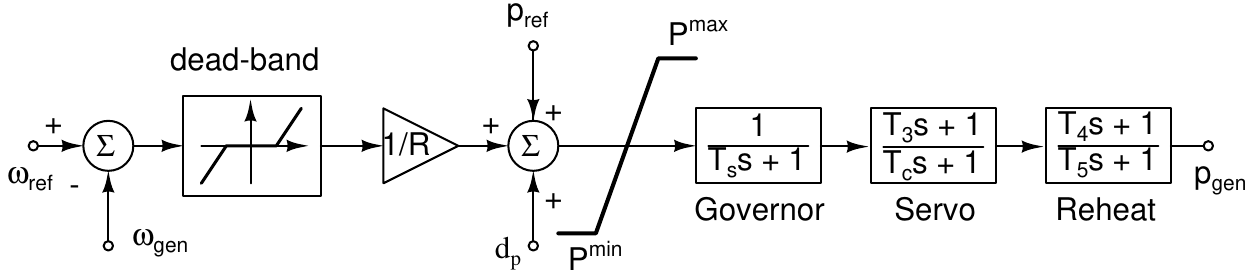}
\caption{Schematic of a \ac{tg} with dead-band.
The $d_{\mathrm{za}}$ value corresponds to the frequency threshold above
which the \ac{tg} modifies the generator power output.
the state space model that implements the transfer function of the
\ac{tg} can be found in \cite{Vancouver}.
\label{F:TgI}}
\end{figure}
%%%%%%%%%%%%%%% END FIGURE %%%%%%%%%%%%%%%
Dead-bands are enabled/disabled according to the specific features that simulations aim at highlighting.
As it is well known, dead-bands are implemented to prevent \acp{tg} from relentlessly operating even following small power variations. In principle, this insertion should not degrade the almost-Gaussian distribution of frequency around the nominal value.
\item
An additional stochastic load is connected to \textsc{bus13}.
It is modeled as
$\displaystyle{p_{\mathrm{L}}(t) = \eta(t) \, P_{\mathrm{L0}}
\left( \frac{\left|v(t)\right|}{V_0} \right)^{\gamma}}\rule{0mm}{7mm}$
\cite{6547228,Vancouver},
where $P_{\mathrm{L0}}$ is the nominal active power of the load,
$V_0$ is the load voltage rating,
$v(t)$ is the bus voltage
%(in the dq-frame)
at which the load is connected,
$\gamma$ governs the dependence of the load
on bus voltage \cite{Vancouver}.
Load active power is ruled by the $\eta(t)$ Ornstein-Uhlenbeck's
process in \Refe{E:Oup} with
${\eta(0)=0}$,
${\alpha = 0.5}$,
${b = 1}$,
${\mu = 0}$ \cite{6547228}\footnote{
Typically, the $\alpha$ parameter ranges from
$\nicefrac{1}{20}$ to $\nicefrac{1}{2}$ \cite{8626538,6547228}.}.
This choice leads to a stochastic load that on average does
not absorb active power.
The purpose of this setup is to stimulate the power system and its controllers
with a continuous active power disturbance.
Two different values of $P_{\mathrm{L0}}$,
namely $1\,\mega\watt$ and $10\,\mega\watt$, are considered.
\item
The conventional load of the \textsc{ieee 14 bus} system
connected at \textsc{bus13}, which has
$P_{\mathrm{D0}} = 14.9\,\mega\watt$ nominal power, is replaced
by the daily time-varying load\footnote{
A similar version of power drifting was already used in \cite{7540970}.}
\begin{equation*}
\displaystyle{p_{\mathrm{D}}(t) = P_{\mathrm{D0}}
    \left( 1 + \Delta p_{\mathrm{D}}(t) \right)
    \left( \frac{\left|v(t)\right|}{V_0} \right)^{\gamma}}~,
\end{equation*}		
where
\begin{equation}
\Delta p_{\mathrm{D}}(t) = -0.12 \sin\left( \frac{t}{24 \times 3600}\right)
~\pu~.
\label{F:VarLoad}
\end{equation}
In the conventional \textsc{ieee 14 bus} without dead-bands,
the addition of this daily
time-varying load causes the corresponding slow intervention of the \acp{tg}
to balance the power mismatch.
Frequency deviates from its nominal value of a very small amount due to the overall low power variation of the drifting load. As shown in the following, the effect of this drifting load on frequency distribution significantly changes when dead-bands are introduced.
\item
To compensate the slowly daily drifting power introduced above, we add
a simple \ac{agc} model as described in \cite{Kundur:1994,7540970}.
The \ac{tg} of generator \textsc{g2} in the \textsc{ieee 14 bus} test
system is driven by a simple \ac{agc} described by the equation
$\dot P_{\mathrm{agc}} = k_{\mathrm{agc}}
\left(1 - \omega_{\mathrm{acg}} \right)$.
In our implementation, $\omega_{\mathrm{acg}}$ is the angular frequency of
the \ac{coi} and the \ac{agc} output $P_{\mathrm{agc}}$ drives the
${\tt d_p}$ terminal of the summation block of the \ac{tg} schematic
shown in \RefFig{F:TgI}.

The choice of adequate \ac{agc} parameters and \acp{tg} dead-bands
to avoid wear and tear may be a difficult task \cite{8727917}.
In any case, regardless of dead-bands, the action of the \ac{agc}
should be very slow and immune from fast frequency fluctuations due
to stochastic loads with relatively short reversal time as in
our case.
\end{itemize}

In every simulated scenario, we always performed an
eigenvalue analysis after the computation of the power-flow solution
to assess whether the modified system remained stable despite
the additions above \cite{Vancouver,Milano:2010,Kundur:1994}.
We also assume that the modified \textsc{ieee 14 bus} power system
is ergodic, stationary, or cyclo-stationary.
The Ornstein-Uhlenbeck's process can be assumed stationary or cyclo-stationary
if we consider its behavior for $\alpha t \gg 0$, as we do in all
scenarios.
Ergodicity allows us to substitute a set of relatively short time-domain
simulations with a single long-lasting time-domain simulation to
derive statistical properties.
The samples of the $\eta(t)$ random variables were generated
according to the mean and variance of the Ornstein-Uhlenbeck Gaussian
process. $\eta(t)$ is thus transformed in a
discrete box-car function on the $1\,\second$ evenly spaced time
grid. Since the $\eta(t)$ samples can be computed
before starting the simulation, we solved a random
differential equation.

In the various scenarios, we simulate the \textsc{ieee 14 bus} power
system for at least 24 hours after a sufficiently large setup time interval.
We pick a frequency sample every $1\,\second$,\footnote{{The sampling
frequency is consistent with that used in the collecting frequency
samples of the Great Britain grid.}}
thereby collecting
 $3600 \times 24 = 86400$ samples of the frequency
variations of the rotor speed of synchronous generators.

\section{Frequency deviations}
\label{S:FreqDev}
In this section, we consider several scenarios with different settings of the
main system's parameters, namely the $H$ inertia of the synchronous generators,
the ${D_{\mathrm{L}}}$ load damping, and dead-bands.
Each subsection considers a peculiar scenario, and we display the obtained
results, i.e., frequency deviations, through histograms.
Our target is to show, through this guided sequence of tailored simulations
of a complex and general system like the \textsc{ieee 14 bus} one,
how the link between $H$ and ${D_{\mathrm{L}}}$ in \Refe{E:Std0}
and \Refe{E:Dl0H} can be exploited to understand their effects on
frequency deviations and possibly how to limit frequency deviations.
Table \ref{T:SetUp} gives a synoptic view of the \textsc{ieee 14 bus}
system setups, used to derive all the histograms in the paper.
\begin{table}[h]
\centering
\caption{Synoptic setup of the histograms. Symbol \bb~means selected.
\label{T:SetUp}}
\includegraphics[width=\columnwidth]{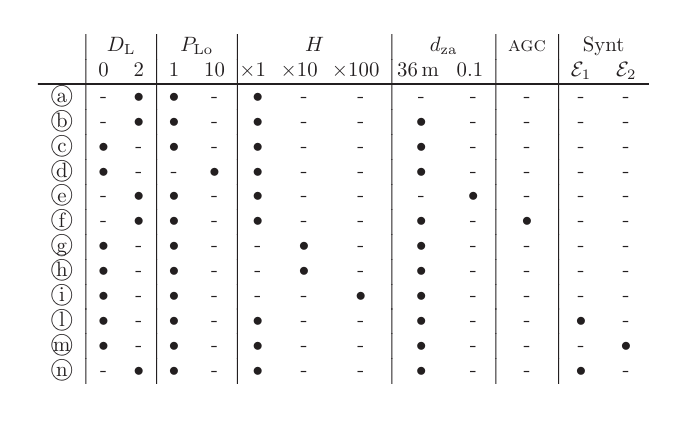}
\end{table}

\begin{figure}[h]
\centering
\includegraphics[width=\columnwidth]{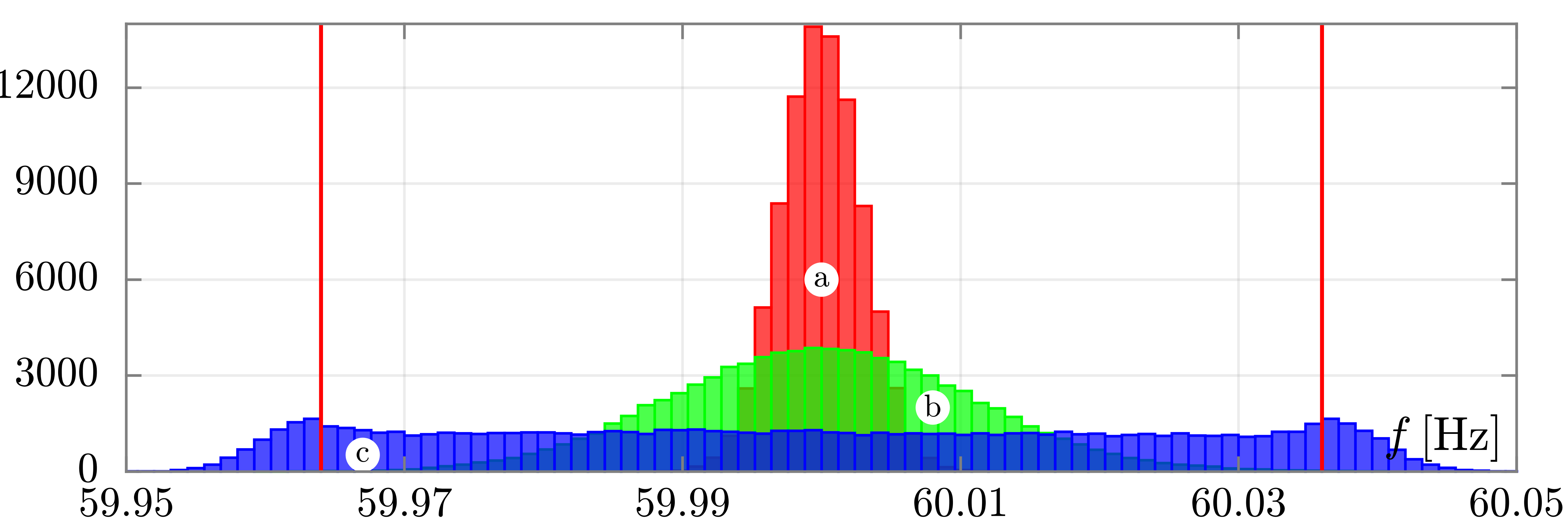}
\caption{The \circled{a} histogram shows the frequency
deviations at nominal conditions and parameters of the \textsc{ieee 14 bus}
power system.
The {\circled{b}} histogram refers to the case with dead-bands and
the {\circled{c}} one refers to the case with dead-bands and with
$D_{\mathrm{L}}=0$.
Vertical lines show the $\mathrm{d_{za}}$ frequency limits of
the dead-band (see \RefFig{F:TgI}).
All histograms are displayed with an equal number of bins (same width).
Histograms report the number of occurrences in each bin.
The value of $P_{\mathrm{L0}}$ is $1\,\mega\watt$.
\label{F:Histo1}}
\end{figure}

To have a reference scenario for frequency deviations and to allow
further easy comparisons, we firstly simulate the nominal \textsc{ieee 14 bus} system by
adding the stochastic load only.
The histogram of frequency deviations obtained in this case is
reported in \RefFig{F:Histo1} [\circled{a}].
%%%%%%%%%%%%%%% BEGIN FIGURE %%%%%%%%%%%%%%%
%%%%%%%%%%%%%%% END FIGURE %%%%%%%%%%%%%%%
The continuous intervention of the \acp{tg}, regulating the mechanical
power from the prime movers that drive the two synchronous generators, leads to a
quite limited frequency variation, which is also due to the modest variance of the
stochastic load.
The resulting frequency histogram is Gaussian.
The frequency deviations with $P_{\mathrm{L0}}=10\,\mega\watt$ are still
limited and are thus not reported.
However, these settings have the disadvantage of requiring the continuous
intervention of \acp{tg}, which is undesirable.

\subsection{\ac{tg} with dead-bands}
We set up a new scenario that adds dead-bands to \acp{tg} by setting
$d_{\mathrm{za}} = 720\times10^{-6}\,\pu$ ($36\,\milli\hertz$) in the
block shown in the schematic of \RefFig{F:TgI}.
The dead-band equally extends of the $d_{\mathrm{za}}$ amount around the
center frequency.
We chose this value since it is compliant with the standards in
\cite{NERC:2012}, i.e., it has practical ground.
We simulate the power system for 24 hours. In the resulting frequency histogram
(see the \circled{b} histogram in \RefFig{F:Histo1}), almost all of the frequency deviation
occurrences lay inside the dead-band, whose thresholds are identified
by vertical solid red lines in \RefFig{F:Histo1}.
It can be seen that in a complex power system there is a
relation in limiting frequency variations linking the variance of the
stochastic model of the loads,
the extension of the dead-bands, and the fraction of \acp{tg} equipped
with dead-bands with respect to those without dead-bands.

An interesting result is the \circled{c} histogram shown in
\RefFig{F:Histo1}.
It is obtained with dead-bands and with ${D_\mathrm{L}=0}$, i.e.,
by fictitiously assuming no dependence of the load power from frequency
as if, for example, each induction motor were equipped with an electronic
controller that decouples it from the grid,
i.e., it keeps its rotor speed independent from grid frequency.
We see an almost flat portion of the histogram inside the dead-band
as predicted by \Refe{E:Std} and \Refe{E:Dl0H}.
There is a modest increase of occurrences just outside the dead-band
where the power fluctuations activate the \acp{tg}.
When frequency exceeds the dead-band, \acp{tg}
activate and pull frequency back.
We recall that the power-flow solution in normal operating conditions
gives a total active power absorbed by loads equal to $259\,\mega\watt$.
When we set $P_{\mathrm{L0}}=1\,\mega\watt$, this value is
about $0.4\%$ of the total active load power.
This power variation, albeit small, leads to frequency fluctuations that are sufficiently large to trigger the intervention of \acp{tg}.
Since we use a low-pass filtered white Gaussian source with variance given
by \Refe{E:OuVar} (recall that we set $b=1$ and $\alpha=0.5$) we tested
the stability of the \textsc{ieee 14 bus} power system with a
maximum power variation of $\pm 5 P_{\mathrm{L0}} = \pm 5 \times 10\mega\watt$.
This corresponds to the
$\mathrm{erf}\left(\frac{5}{\sqrt{2}}\right) \simeq 1 - 5.7330 \times 10^{-7}$
expected fraction of the generated samples that fall inside the
$\pm 5 P_{\mathrm{L0}}$ interval.
We expect that the \textsc{ieee 14 bus} power system
almost always works in a stable condition.
This removes the suspect that the frequency deviations are due to the system being unstable.
The first consideration about the \circled{c} histogram is that
{the complete elimination
of any dependence of the power from frequency, i.e., the setting of
${D_\mathrm{L}=0}$, defeats the purpose of inserting dead-bands to limit
wear and tear of \acp{tg}}.
This also largely justifies our choice of frequency dead-bands.
We stress that an unaware and independent
selection of the dead-band settings from the aggregate $D_\mathrm{L}$
value of a power system might be useless.
The second consideration is that frequency distribution greatly
deteriorates.
We see that the shape of the histogram largely differs from a Gaussian one
since it is practically flat inside the dead-band.
This means that frequency {\em uniformly} fluctuates inside the dead-band.
We performed several simulations by progressively lowering $D_{\mathrm{L}}$
from 2 to 0.
%%%%%%%%%%%%%%% BEGIN FIGURE %%%%%%%%%%%%%%%
\begin{figure}[t]
\centering
\includegraphics[width=\columnwidth]{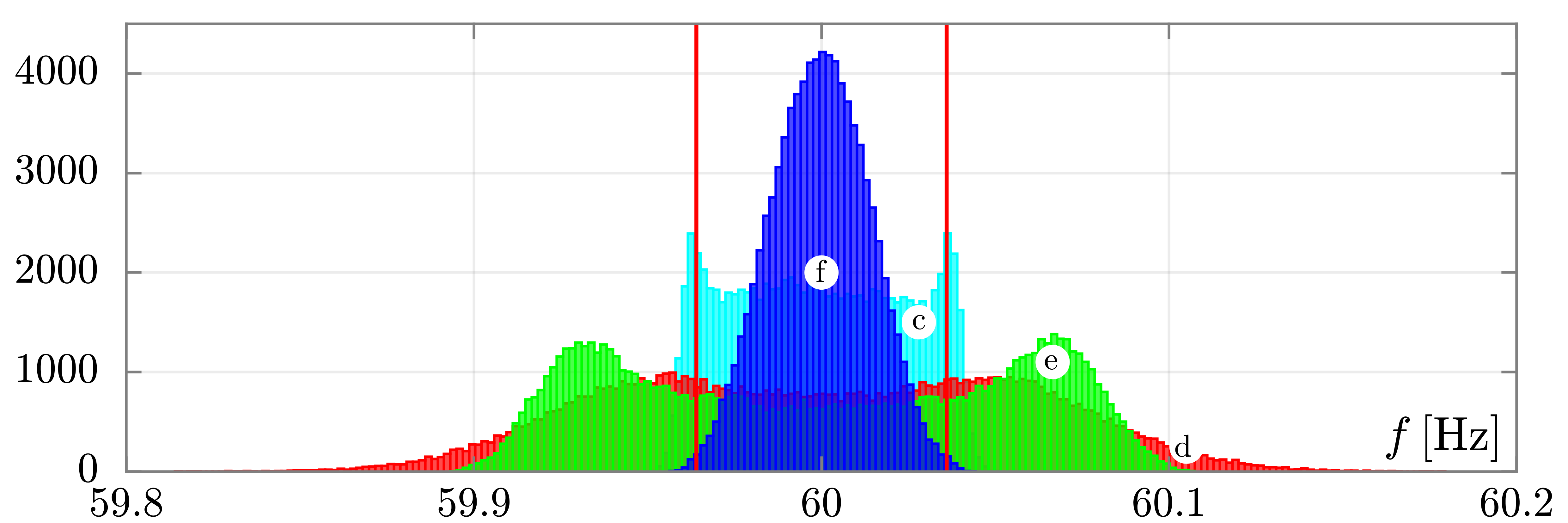}
\caption{
Frequency histograms.
{\circled{c}} histogram: $\mathrm{d_{za}} = 36\,\milli\hertz$,
$D_{\mathrm{L}}=0$ and $P_{\mathrm{L0}}$ is $1\,\mega\watt$.
\circled{d} histogram: $\mathrm{d_{za}} = 36\,\milli\hertz$,
$D_{\mathrm{L}}=0$ and $P_{\mathrm{L0}}$ is $10\,\mega\watt$.
\circled{e} histogram: $D_{\mathrm{L}} = 2.0$,
$\mathrm{d_{za}} = 100\,\milli\hertz$
$P_{\mathrm{L0}}=1\,\mega\watt$, power drifting is introduced
(see \RefEq{F:VarLoad}).
The \circled{c}, \circled{d}, and \circled{e} histograms are bimodal.
The \circled{f} histogram refers to the same scenario of the
\circled{e} one but with the introduction of the \ac{agc}
($k_{\mathrm{agc}}=0.01$)
that drives the \ac{tg} of the \textsc{g2} generator and
with $\mathrm{d_{za}} = 36\,\milli\hertz$.
Vertical lines mark the extension of the dead-band corresponding to the \circled{d} and \circled{f} histograms.
\label{F:Histo2}}
\end{figure}
%%%%%%%%%%%%%%% END FIGURE %%%%%%%%%%%%%%%

We do not report them due to space reasons.
What we obtained is that by starting from the \circled{b} histogram,
the other histograms progressively flatten as $D_{\mathrm{L}}$ lowers
till becoming the \circled{c} histogram.
In doing this we observed a progressive increase of frequency
occurrences outside the dead-band, i.e. interventions of \acp{tg}.
Coherently with \Refe{E:Std}, we also state that {frequency deterioration
cannot be effectively compensated by acting on the $H$
inertia of the system if $D_{\mathrm{L}}$ is too low.
This deterioration further worsens as the stochastic load variation increases}.
To enforce this statement, we repeat the last simulation with a higher ${P_{\mathrm{L0}}}$ value ($10\,\mega\watt$), corresponding to about $4\%$ of the total nominal load power.
We expect larger occurrences of frequency deviations outside the dead-bands
and mainly more interventions of \acp{tg}.
The \circled{d} frequency histogram, corresponding to this scenario, is
shown in \RefFig{F:Histo2} in which the \circled{c} histogram of
\RefFig{F:Histo1} is reported too to facilitate comparisons.
It highlights the presence of a {\em bimodal} frequency distribution
\cite{7540970}, characterised by two symmetric peaks of frequency
occurrences located just outside the dead-bands, where \acp{tg} start
to counteract frequency deviations.
Such occurrences increase with the variance of the stochastic load. Therefore, an adequately large variance might result in bimodal frequency histograms even if
$D_{\mathrm{L}} \gg 0$.

\subsection{Bimodal histograms}
In computing the results of previous scenarios we have assumed that dead-bands
equally extend from the $60\,\hertz$ center frequency.
Said oppositely, we have assumed that the average value of
frequency is $60\,\hertz$ and thus it locates exactly in the mid-point
of the dead-band.
We now introduce the deterministic load power drift described by
\Refe{F:VarLoad}. We assume that the load at \textsc{bus13}
slowly varies during the 24 hours along which we simulate the
\textsc{ieee 14 bus} power system.
This power imbalance causes a corresponding frequency shift.

\begin{figure}[h]
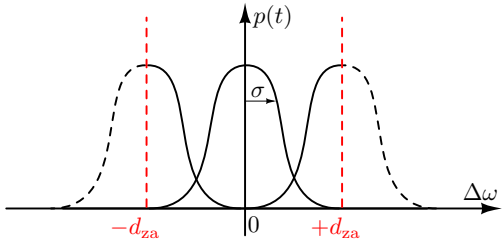

\centering
\include{./cycle}
\caption{Qualitative effect of the daily time-varying load on frequency distribution when \ac{agc} is absent ({\em p(t)} denotes probability).}
\label{F:Cycle}
\end{figure}

By considering our simplified power system model and the daily time-varying load, \Refe{E:Mean2} and \Refe{E:Std} show that the average value of $\Delta \omega$ periodically cycles around 0 with a constant $\sigma$ variance, i.e., the central Gaussian curve in \RefFig{F:Cycle} shifts horizontally back and forth over time. If the average frequency variation and $\sigma$ are large enough, we expect that the probability of crossing the dead-band increases cyclically. The result will be a bimodal histogram with equal peaks of occurrences outside of the dead-band.
%%%%%%%%%%%%%%% BEGIN FIGURE %%%%%%%%%%%%%%%
%%%%%%%%%%%%%%% END FIGURE %%%%%%%%%%%%%%%
We simulated the system with ${D_{\mathrm{L}} = 2}$,
${\mathrm{d_{za}} = 100\,\milli\hertz}$, i.e., with a larger dead-band,
and ${P_{\mathrm{L0}}=1\,\mega\watt}$, thereby obtaining the \circled{e}
frequency histogram shown in \RefFig{F:Histo2}.
This new choice of dead-band is coherent with values by
other \textsc{tso}s.
Note that this large value should better contribute to keeping \acp{tg}
inactive and thus to obtain a well-shaped (Gaussian) frequency histogram.

It can be noticed that no \ac{agc} was introduced to compensate for the slowly
varying deterministic power.
As a consequence, the average frequency value varies in a slow (quasi-static) manner above and below $60\,\hertz$.
It translates into going closer to the upper and lower limits of the dead-band. Thus, as already stated, the likelihood of crossing the dead-band and
triggering the activation of \acp{tg} increases,
even though ${D_{\mathrm{L}} = 2}$.

We underline that the \circled{d} and \circled{e} histograms
are obtained in two different scenarios.
In the former, we have a too low (i.e., ${D_{\mathrm{L}}=0}$) load damping
and, in the latter, we have a poorly managed, slowly varying power imbalance.
Although the power system characteristics and setups of the two scenarios are very different, they both result in bimodal histograms.
%%%%%%%%%%%%%%% BEGIN FIGURE %%%%%%%%%%%%%%%
\begin{figure}[h]
\centering
\includegraphics[width=\columnwidth]{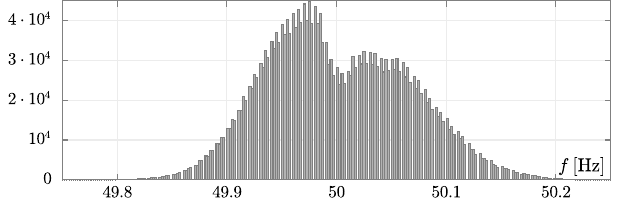}
\caption{Frequency histogram of the Great Britain grid, June 2018.
The histogram bins are uniformly distributed in the frequency interval $[49.75, 50.25]\,\hertz$
and the bin amplitude is $2\,\milli\hertz$.
\label{F:GbhJune}}
\end{figure}
%%%%%%%%%%%%%%% END FIGURE %%%%%%%%%%%%%%%

%%%%%%%%%%%%%%% BEGIN FIGURE %%%%%%%%%%%%%%%
\begin{figure*}[t]
\centering
\includegraphics[width=\textwidth]{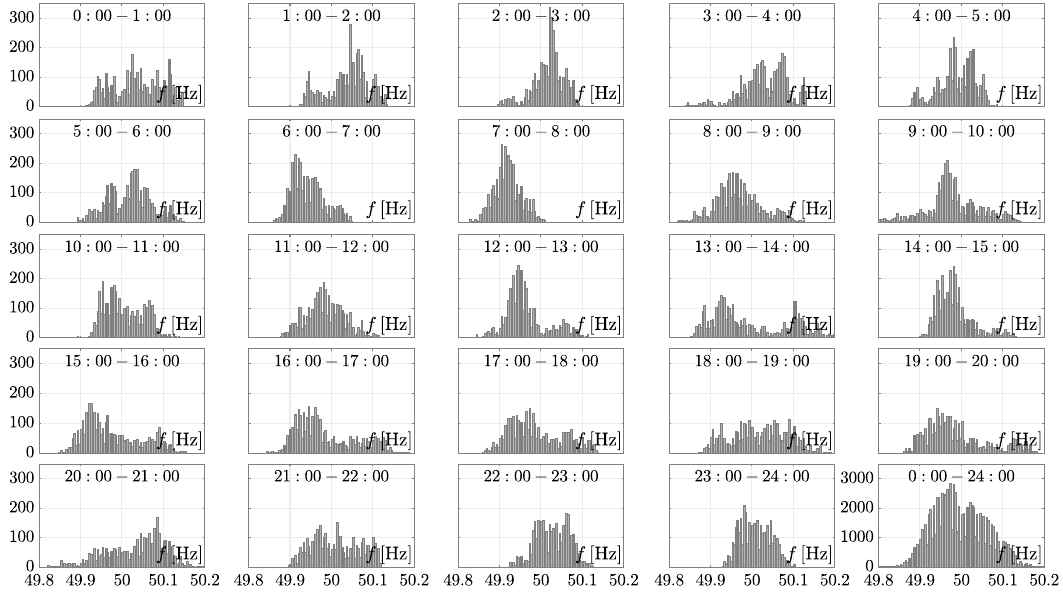}
\caption{Frequency histograms of the Great Britain grid, June $\text{1}^{\text{st}}$, 2018. Histograms display occurrences of frequency deviations on an hourly basis. The histogram in the lower-right corner considers the samples of the entire day. The histogram bins are uniformly distributed in the frequency interval $[49.8, 50.2]\,\hertz$ and the bin amplitude is $4\,\milli\hertz$.
\label{F:GbhJune1}}
\end{figure*}
%%%%%%%%%%%%%%% END FIGURE %%%%%%%%%%%%%%%

\subsection{Bimodal frequency histograms of the Great Britain's grid}
Let us now consider a real situation, namely the frequency histogram
of the Great Britain grid displayed from data recorded along with June 2018.
The frequency samples are of the public domain and provided by the
{\em National Grid ESO}, which is the electricity system operator for Great Britain \cite{GB:2016}.
The historic frequency data for Great Britain are provided at a $1 \,\second$ resolution,
as in our simulations. The related histogram is depicted in \RefFig{F:GbhJune}. It refers to the {entire month of June}. It can be seen that it has a bimodal shape. Our rhetorical question is: ``Which kind of bimodal histogram is it?'' Better said: ``Is it of type \circled{d} (too small $D_{\mathrm{L}}$) or \circled{e} (inadequately handled daily power imbalance)?'' The answer comes if we observe the hourly frequency histograms of the Great Britain grid for example on June $\text{1}^{\text{st}}$, 2018.

Analogous results are obtained by considering other days.
Each hourly histogram seems {\em unimodal}, and none of them clearly shows
any neat bimodal shape.
The histograms shown in \RefFig{F:GbhJune1} highlight that there is a
daily frequency drift possibly due to a mismatch of power
demand and planned generation not adequately corrected by \acp{agc} and that
there is some sort of dead-band.
The shape of the hourly and daily histograms suggest that
there may be an adequate level of load damping ($D_{\mathrm{L}}$) introduced
by frequency-dependent loads.

In the monthly histogram of \RefFig{F:GbhJune} there is an excess of
occurrences due to a lack of power production with respect to
power demand that leads to the asymmetry of the two peaks
below and above $50\,\hertz$.
The number of occurrences below $50\,\hertz$ exceeds those above.
We are aware that our statements are a long shot at best since we do not
have detailed information about the characteristics of the Great Britain grid.
However, we dare say that monthly frequency samples aggregate may not
constitute a valid indicator of the true nature (i.e., unimodal or bimodal)
of power system frequency.
Thus, conclusions drawn ``tout court'' from those could be misguided.
For instance, in the case of Great Britain's grid, a seemingly bimodal
monthly frequency histogram could be due to the sum of unimodal hourly
histograms collected in a power system prone to daily frequency drift.

The correct aggregation window for frequency samples depends on power
system characteristics.
In our scenarios, an hourly window should be adequate, since it is a time interval large enough with respect to typical power system time constants to collect a sufficient number of samples (ergodic process), but short enough compared to the slow drifting load dynamics.
The same seems true for Great Britain's case\footnote{ One of the anonymous reviewers gave us the good suggestion that the bimodal shape of the Great Britain's frequency histogram: ``... could be related to the
Frequency And Time Error (\textsc{fate}) mechanism used by National Grid to align grid frequency with time, requiring operation for periods slightly above or below the setpoint.'' The mentioned ``periods'' are time intervals that may last for several hours till one or two days, thus well above the proposed hourly aggregation time to derive histograms. The new setpoint introduces frequency deviations of only tens of
$\milli\hertz$ \cite{Fate}.}.

Another aspect worth further discussing is that the insertion
of dead-bands does not limit wear and tear of \acp{tg} if the frequency
drifting is not properly compensated and the frequency equilibrium point
(short term average frequency) is not kept as close as possible to the
nominal value.

\subsection{Automatic generation control}
The frequency drifting caused by the daily variation of the
load/generation imbalances can be compensated by a well designed \ac{agc}. In order to address this issue, we insert in the \textsc{ieee 14 bus} power system  the simple \ac{agc} model described in Section \ref{S:NumSetUp}.
The transfer function of the \ac{agc} is made up of a single pole in the frequency origin. Its gain has to be set so that it counteracts the hourly frequency variation caused by the deterministic drifting load, while practically filtering out the fast fluctuations due to the stochastic load.

We repeated the
simulation of the scenario with the daily variable load
but with narrower dead-bands.
The result obtained by simulating this scenario with the
$D_{\mathrm{L}} = 2.0$, $\mathrm{d_{za}} = 36\,\milli\hertz$ and
$P_{\mathrm{L0}}=1\,\mega\watt$ parameters is given by the \circled{f} histogram depicted in \RefFig{F:Histo2}.
It is immediate to appreciate that the \ac{agc} largely compensates
for the deterministic power drift.
The \circled{f} histogram is completely different from the other two in the same figure and resembles the shape of a Gaussian
probability density function that adequately lays inside the dead-band.

This result has suggested our comments on the daily histograms of the
Great Britain's grid, i.e., that there may be a poor compensation of even small
power imbalances which leads to a bimodal monthly frequency
histogram\footnote{In \cite{6855377} the authors state at the beginning of
column 2, page 702: ``It is important to note that, ....., there is no
automatic generation control \ac{agc} implemented on the Great
Britain's grid''.
Some more information on dead-bands and on the load-frequency control process,
which has the purpose of addressing imbalances close to real time through the
sequential activation of control reserves, is reported in \cite{8469998}.
The frequency control and restoration reserves are automatically activated
with the purpose of stabilizing frequency around a new setpoint value.
In \cite{8469998} there is a list of references from which data
were derived.}.

\section{Inertia}
\label{S:Inertia}
In presenting the frequency histograms obtained in the scenarios considered so far, we have not explicitly paid attention to the system inertia.
Better said, we kept fixed the $H$ inertia parameter of all the synchronous
generators and compensators of the \textsc{ieee 14 bus} power
system.
In \cite{8626538} it is stated that the main factors influencing
frequency distribution are the load damping and
\ac{tg} dead-bands, while inertia is reputed to
have a minimal impact. Actually, according to \Refe{E:Std0}, inertia
positively influences frequency deviations, i.e.,
the denominator is greater than 1 and  $\sigma < b$,
if
${H \gg \nicefrac{(1-2\alpha D_{\mathrm{L}}^2)}{(4 \alpha^2 D_{\mathrm{L}})}}$.
We remark that if $D_{\mathrm{L}}^2$ is sufficiently large the right hand-side
of the inequality becomes negative.
It means that $H$ has a marginal role in limiting frequency
deviations.
Once more, we underline that $D_{\mathrm{L}}$ must assume suitable values
with a proper level of $H$,
otherwise the Gaussian shape of the probability density function flattens
as highlighted by the simulated scenarios of Section \ref{S:FreqDev}.
It indicates that if the inertia of a power system decreases
due to higher penetration of \ac{res}-fuelled converter-connected generation
(and converter-connected loads), the frequency deviation can be still kept
small if the value of $D_{\mathrm{L}}$ is adequately large.
For example, large $D_{\mathrm{L}}$ values can be obtained by
``thermostatically'' controlling load such as fridges, air conditioners, and
boilers \cite{CONTE2017291}.

To show the influence of inertia on frequency deviations, we considered
three different scenarios.
In each of these scenarios we have ${\mathrm{d_{za}} = 36\,\milli\hertz}$,
${P_{\mathrm{L0}}=1\,\mega\watt}$ and ${D_{\mathrm{L}} = 0}$.
Furthermore, we do not introduce any daily time-varying load, and thus there is no need to use \acp{agc}.

In the first and second scenarios, we multiply the $H$ inertia of
each synchronous generator and compensator by ten times.
Having in mind \Refe{E:Dl0H} and \Refe{E:Dl0H2}, in the first scenario, the
simulation lasts one day, whereas in the second scenario it lasts four days.
In the third scenario, we multiply by $100$ the inertia, thus
remarkably increasing it and collect samples along eight days.
The increase of the simulation time interval is dictated by the fact
that $H^2$ is at the denominator in \Refe{E:Dl0H}.

\begin{figure}[h]
\centering
\includegraphics[width=\columnwidth]{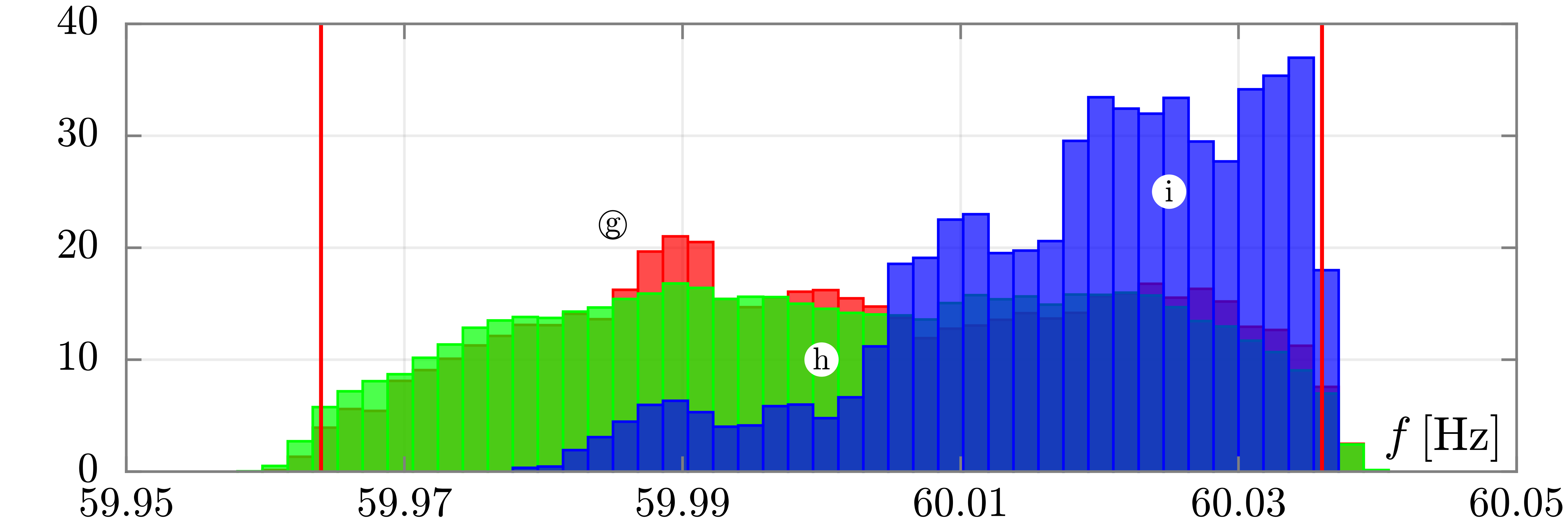}
\caption{The \circled{g} ($10 \times H$, 1 day),
\circled{h} ($10 \times H$, 4 days) and
\circled{i} ($100 \times H$, 8 days)
histograms correspond to the first, second and third scenarios, respectively.
$\mathrm{d_{za}} = 36\,\milli\hertz$, $P_{\mathrm{L0}}=1\,\mega\watt$
$D_{\mathrm{L}} = 0$.
The histograms are normalised to estimate the probability density.
\label{F:Histo3}}
\end{figure}

According to the common belief that inertia restoration is a must
against frequency degradation, we would expect very limited
frequency deviations and thus a very well shaped histogram.
The three histograms corresponding to these scenarios are shown in
\RefFig{F:Histo3}.
%%%%%%%%%%%%%%% BEGIN FIGURE %%%%%%%%%%%%%%%
%%%%%%%%%%%%%%% END FIGURE %%%%%%%%%%%%%%%
They are normalised to estimate the probability density
and, the bins do not count occurrences as in the previous histograms.
It is done since the number of samples largely
differs in these three scenarios.
The main aspect that we easily see is that frequency deviates from
$60\,\hertz$ with almost the same probability inside the dead-band.
When the frequency exits the dead-bands, the \acp{tg} intervene to
pull the frequency back inside the dead-band.
They constitute some sort of ``barrier''.
These histograms do not have at all the shape of a Gaussian
centered at $60\,\hertz$, characterised by a very small variance
despite the very large increase of system inertia.

We believe that this is a relevant result, which is in contrast to the
common belief that inertia must be increased independently from
any other aggregate parameter of a power system to counteract the
negative effects of larger penetration of \ac{res} generation.

\section{Synthetic inertia}
\label{S:SynIn}
Most of \ac{res} can be regarded as stochastic sources due to the volatility
of their primary energy source (e.g., wind and sun). It means that the
effects of their power output variability on frequency
can be considered very similar to those of the stochastic load described in
the previous sections, namely, an increase of generation can be compared to a
decrease in load demand and vice versa.
The main difference may be related to the statistical properties of these
sources.

\ac{res} can be generally considered inertia-less for two main reasons.
First, as in the case of solar plants, they lack a mechanical
rotating mass, which makes them incapable of exchanging and converting kinetic
energy after a power mismatch to limit frequency variations.
Second, unless synthetic inertia is implemented, such plants are
typically connected to the grid through power electronic converters
whose control schemes do not allow to modify their power output based
on frequency fluctuations.

In this section, we evaluate how frequency variations can be mitigated
through the installation of an aggregated \ac{pv} power plant connected
to the \textsc{ieee 14 bus} power system at $\textsc{bus11}$ through
a converter implementing \textit{synthetic inertia}.
The word \textit{synthetic} means that the converter controls are such
that, when a power mismatch occurs, the behavior of the power plant in
terms of power and energy exchange resembles that of a synchronous generator.
It is worth noticing that synthetic inertia does an ``endogenous'' action by leveling
the fluctuations of the \ac{res} and an ``exogenous'' action by
counterbalancing the stochastic variations of external loads.

\begin{figure}[h]
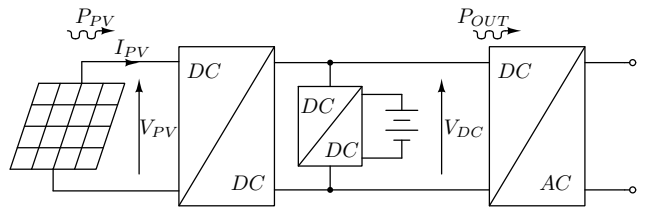

\begin{center}
\include{./pvschemeBat}
\caption{Connection scheme of the \ac{pv} plant and battery pack to the \textsc{ieee 14 bus} power system. $V_{\mathrm{pv}}^{\mathrm{ref}} = 700 \volt$, $I_{\mathrm{pv}}^{\mathrm{ref}} = 2.8571 \kilo\ampere$, $V_{\mathrm{dc}}^{\mathrm{ref}} = 680 \volt$. The superscript $^\mathrm{ref}$ denotes the reference \ac{pv} plant operating conditions considered during simulation.}
\label{fig:PVschemeBat}
\end{center}
\end{figure}

The block schematic of the \ac{pv} power plant with synthetic inertia is  shown in \RefFig{fig:PVschemeBat}.
%%%%%%%%%%%%%%% BEGIN FIGURE %%%%%%%%%%%%%%%
%%%%%%%%%%%%%%% END FIGURE %%%%%%%%%%%%%%%
In this design, we are only interested in the high-level behavior of the
system and thus we acted with this target during its design.
A \ac{pv} field supplies power to the \textsc{dc/dc} converter that is equipped
with a maximum power point tracker.
The \textsc{dc} link voltage is kept constant by the \textsc{dc/dc}
bidirectional converter connected to the battery pack.
The power delivered/absorbed by the bidirectional \textsc{dc/ac} converter
is set by the maximum power point tracker, by the \textsc{dc/dc}
converter and by the specific control schema that implements synthetic inertia.
An important aspect of implementing the synthetic inertia controller is
the maximum power that can be delivered by the \textsc{dc/ac} converter and
the energy capacity of the storage system.
In our case, the peak power capacity of the \ac{pv} field and the
\textsc{dc/ac} converter are respectively $1$ and $2\,\mega\watt$, whereas
the energy stored in the battery pack is
${\cal E}_1=160\,\mega\joule$
or ${\cal E}_2=320\,\mega\joule$.
These design choices allow for power and energy variations comparable
to those of the stochastic load.
As a consequence, on average synthetic inertia does not contribute either
net power or energy.

We performed simulations with the addition of synthetic inertia
and with different setups.
In the first case, we set $D_{\mathrm{L}}=0$, $P_{\mathrm{L0}}=1\,\mega\watt$ and
used the ${\cal E}_1$ battery pack; in the second case, we used
the ${\cal E}_2$ battery pack, and in the third case, we used
${\cal E}_1$ but set $D_{\mathrm{L}}=2$.
In all these cases, we set $\mathrm{d_{za}}=36\,\milli\hertz$ and turned off
the daily time-varying load and the \ac{agc}.

\begin{figure}[h]
\centering
\includegraphics[width=\columnwidth]{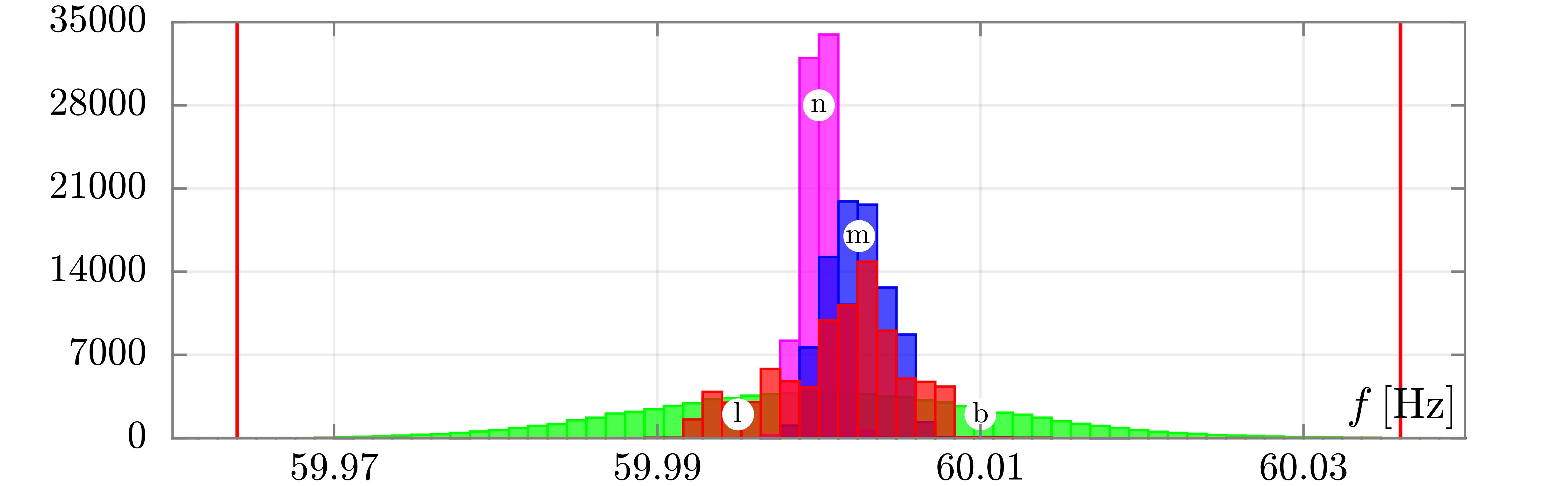}
\caption{The frequency histograms obtained when the \ac{pv} field
implementing synthetic inertia is added to the \textsc{ieee14} power system
with no \ac{agc}, no daily time-varying load, and $P_{\mathrm{L0}}=1\,\mega\watt$.
Vertical lines correspond to the   dead-band thresholds. $D_{\mathrm{L}}=0$
in the \circled{l} (${\cal E}_1$ battery pack) and \circled{m} (${\cal E}_2$ battery pack) histograms. $D_{\mathrm{L}}=2$
in the \circled{n} (${\cal E}_1$ battery pack).
The \circled{b} histogram is the same of \RefFig{F:Histo1}.
\label{F:Histo4}}
\end{figure}

The results obtained in the three cases are depicted in \RefFig{F:Histo4}.
We replicated the {\circled{b}} histogram of \RefFig{F:Histo1} in
\RefFig{F:Histo4}.
These histograms can be compared to assess the significant,
beneficial effect of synthetic inertia on the frequency distribution.
%%%%%%%%%%%%%%% BEGIN FIGURE %%%%%%%%%%%%%%%
%%%%%%%%%%%%%%% END FIGURE %%%%%%%%%%%%%%%
The histogram of \RefFig{F:Histo4} shows that when synthetic inertia
is active almost all frequency deviations occurrences concentrate in a
very small interval around $60\,\hertz$.
It happens since the reversal time of the stochastic load is $2\,\second$
while the bandwidth of the controller implementing synthetic inertia
is about $1\,\kilo\hertz$.
The controller is relatively fast and efficiently manages the stochastic
load by practically keeping frequency close to $60\,\hertz$.
When we use the ${\cal E}_2$ battery pack, the frequency deviations
are narrower than those obtained with the ${\cal E}_1$ battery pack.
Both histograms do not have a Gaussian shape since $D_{\mathrm{L}}=0$.
The interesting aspect is that when $D_{\mathrm{L}}=2$ and we use the
${\cal E}_1$ battery pack we obtain a well-shaped Gaussian histogram
with a peak of occurrence greater than that of the {\circled{a}} case
of \RefFig{F:Histo1} that was obtained without dead-bands, i.e., with
the contribution of \acp{tg} to control frequency deviations.
In this case, there was not any intervention by \acp{tg} to control
frequency.
The need for relatively small capacity of the battery pack with respect
to the very large kinetic energy stored in the full \textsc{ieee 14 bus}
system is since it can be largely and quickly exchanged
with the grid even when frequency variation is small.
It does not apply to a synchronous generator since the exchanged
energy is only a fraction of the stored kinetic energy due to the
limited relative frequency variation.
It shows that the use of synthetic inertia may be extremely convenient and
viable to control frequency fluctuations due to stochastic loads
\cite{8626538,TANG2019197,MAGDY2019351}.

\section{Conclusion}
\label{S:Conclusions}\balance
In this paper, we focused on the reasons that lead to unwanted frequency
distributions, such as bimodal ones.
We presented both analytical and numerical results.
The former
derive from the rigorous analysis of a simplified and linear power system model
and allowed us to highlight some important focal points.
\begin{itemize}
\item The smaller the  load damping coefficient, the lower the effectiveness
of inertia in limiting frequency deviations.
This is in contrast to the common idea that an increment of the overall
system inertia
is beneficial, regardless of the value of other aggregate system parameters.
\item  The  load damping coefficient is deemed to play
a significant role in limiting frequency variations and it could be
increased by exploiting for example thermostatically controlled loads that link
frequency variations to the absorbed power, as well as
load damping and synthetic inertia, which require a suitable implementation
of \ac{res} generation and storage systems.
\item The reduction of the  load damping coefficient leads to an increased probability of
crossing the dead-bands implemented in \acp{tg} and, thus, of obtaining bimodal frequency histograms.
\end{itemize}
According to the numerical results that we derived through a set of
simulations of the well-known \textsc{ieee 14 bus} power system,
we draw the following conclusions.
\begin{itemize}
\item
Inertia together with turbine controls plays a significant role
to contain large frequency deviations and thus
to keep operational security, when severe contingencies happen in a
power grid.
However, in case of small stochastic fluctuations of load/generation,
an increase, even extremely large, of system inertia does not lead to
a Gaussian frequency distribution if the load damping does not have a proper
value.
This means that to have small and well-shaped frequency variations
we have to simultaneously act on these two system parameters.
\item Any slow deterministic frequency drift, albeit
small, must be compensated to avoid tear and wear of \acp{tg} when
dead-bands are implemented.
For example, this has to be accomplished by well-designed \acp{agc}.
Acting on electronic controllers of load and \ac{res} is beneficial if they
both increase inertia (synthetic) and load damping (synthetic).
\end{itemize}
These considerations have general validity since our analysis considered
the \textsc{ieee 14 bus} power system as one among several ones.
We could have chosen other known power system models to draw the same conclusions above
 (as we did to test results). The reported histograms of frequency deviations constitute a logic
``path'' through which the reader is conducted to the considerations we
draw.
In the light of the conclusions, a real situation, namely the frequency
histogram of the Great Britain grid displayed from data recorded along with
June 2018, was analyzed.
It led us also to some considerations concerning the definition of a
correct aggregation procedure to deal with historic frequency data.

%------------------------------------------------------------------------------------
\bibliographystyle{IEEEtran}

\end{document}

%% file: cycle.tex
% XCircuit output "cycle.tex" for LaTeX input from cycle.eps
\def\putbox#1#2#3#4{\makebox[0in][l]{\makebox[#1][l]{}\raisebox{\baselineskip}[0in][0in]{\raisebox{#2}[0in][0in]{\scalebox{#3}{#4}}}}}
\def\rightbox#1{\makebox[0in][r]{#1}}
\def\centbox#1{\makebox[0in]{#1}}
\def\topbox#1{\raisebox{-0.60\baselineskip}[0in][0in]{#1}}
\def\midbox#1{\raisebox{-0.20\baselineskip}[0in][0in]{#1}}
   \scalebox{1.5}{
   \normalsize
   \parbox{1.72917in}{
   \includegraphics[scale=0.666667]{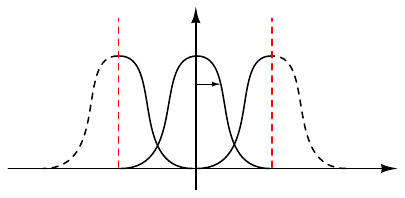}\\
   % translate x=416 y=146 scale 0.56
   \putbox{0.40in}{0.04in}{0.60}{$\red{-d_{\mathrm{za}}}$}%
   \putbox{1.10in}{0.04in}{0.60}{$\red{+d_{\mathrm{za}}}$}%
   \putbox{0.90in}{0.78in}{0.60}{$p(t)$}%
   \putbox{0.88in}{0.05in}{0.60}{$0$}%
   \putbox{0.89in}{0.52in}{0.60}{$\sigma$}%
   \putbox{1.63in}{0.16in}{0.60}{$\Delta \omega$}%
   } % close 'parbox'
   } % close 'scalebox'
   \vspace{-\baselineskip} % this is not necessary, but looks better

%% file: pvschemeBat.tex
% XCircuit output "pvschemeBat.tex" for LaTeX input from pvschemeBat.eps
\def\putbox#1#2#3#4{\makebox[0in][l]{\makebox[#1][l]{}\raisebox{\baselineskip}[0in][0in]{\raisebox{#2}[0in][0in]{\scalebox{#3}{#4}}}}}
\def\rightbox#1{\makebox[0in][r]{#1}}
\def\centbox#1{\makebox[0in]{#1}}
\def\topbox#1{\raisebox{-0.60\baselineskip}[0in][0in]{#1}}
\def\midbox#1{\raisebox{-0.20\baselineskip}[0in][0in]{#1}}
   \scalebox{0.5}{
   \normalsize
   \parbox{6.54688in}{
   \includegraphics[scale=2]{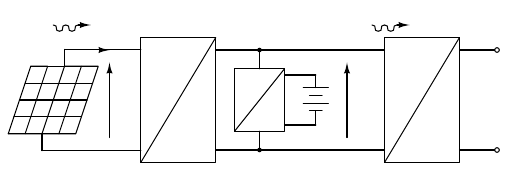}\\
   % translate x=787 y=416 scale 0.19
   \putbox{0.79in}{2.03in}{1.60}{$P_{PV}$}%
   \putbox{1.20in}{1.70in}{1.60}{$I_{PV}$}%
   \putbox{1.96in}{1.45in}{1.60}{$DC$}%
   \putbox{2.42in}{0.28in}{1.60}{$DC$}%
   \putbox{5.21in}{1.45in}{1.60}{$DC$}%
   \putbox{5.65in}{0.28in}{1.60}{$AC$}%
   \putbox{1.46in}{0.86in}{1.60}{$V_{PV}$}%
   \putbox{4.79in}{2.03in}{1.60}{$P_{OUT}$}%
   \putbox{4.65in}{0.86in}{1.60}{$V_{DC}$}%
   \putbox{3.15in}{1.08in}{1.60}{$DC$}%
   \putbox{3.40in}{0.62in}{1.60}{$DC$}%
   } % close 'parbox'
   } % close 'scalebox'
   \vspace{-\baselineskip} % this is not necessary, but looks better

%% file: inertia4arXiv.bbl
\begin{thebibliography}{10}
\providecommand{\url}[1]{#1}
\csname url@samestyle\endcsname
\providecommand{\newblock}{\relax}
\providecommand{\bibinfo}[2]{#2}
\providecommand{\BIBentrySTDinterwordspacing}{\spaceskip=0pt\relax}
\providecommand{\BIBentryALTinterwordstretchfactor}{4}
\providecommand{\BIBentryALTinterwordspacing}{\spaceskip=\fontdimen2\font plus
\BIBentryALTinterwordstretchfactor\fontdimen3\font minus
  \fontdimen4\font\relax}
\providecommand{\BIBforeignlanguage}[2]{{%
\expandafter\ifx\csname l@#1\endcsname\relax
\typeout{** WARNING: IEEEtran.bst: No hyphenation pattern has been}%
\typeout{** loaded for the language `#1'. Using the pattern for}%
\typeout{** the default language instead.}%
\else
\language=\csname l@#1\endcsname
\fi
#2}}
\providecommand{\BIBdecl}{\relax}
\BIBdecl

\bibitem{Kundur:1994}
P.~Kundur, \emph{Power system stability and control}.\hskip 1em plus 0.5em
  minus 0.4em\relax New York: McGraw-Hill, 1994.

\bibitem{NERC:2011}
{\relax North American Electric Reliability Corporation (NERC)},
  \emph{Balancing and Frequency Control}, 2011.

\bibitem{NERC:2012}
------, \emph{Frequency Response Initiative Report: The Reliability Role of
  Frequency Response}, 2012.

\bibitem{Kroposki:2017}
B.~Kroposki, B.~Johnson, Y.~Zhang, V.~Gevorgian, P.~Denholm, B.~. Hodge, and
  B.~Hannegan, ``Achieving a 100\% renewable grid: Operating electric power
  systems with extremely high levels of variable renewable energy,''
  \emph{{IEEE} Power Energy Mag.}, vol.~15, no.~2, pp. 61--73, 2017.

\bibitem{ARRIGO2020105428}
F.~Arrigo, E.~Bompard, M.~Merlo, and F.~Milano, ``Assessment of primary
  frequency control through battery energy storage systems,'' \emph{Int. J.
  Electr. Power Energy Syst.}, vol. 115, p. 105428, 2020.

\bibitem{Obaid:2019}
Z.~A. Obaid, L.~M. Cipcigan, L.~Abrahim, and M.~T. Muhssin, ``Frequency control
  of future power systems: reviewing and evaluating challenges and new control
  methods,'' \emph{J. Mod. Power Syst. Clean Energy}, vol.~7, no.~1, pp. 9--25,
  2019.

\bibitem{Ulbig:2014}
A.~Ulbig, T.~S. Borsche, and G.~Andersson, ``Impact of low rotational inertia
  on power system stability and operation,'' in \emph{IFAC Proceedings Volumes
  (IFAC-PapersOnline)}, vol.~19, 2014, pp. 7290--7297.

\bibitem{Tielens:2016}
P.~Tielens and D.~Van~Hertem, ``The relevance of inertia in power systems,''
  \emph{Renew. Sust. Energ. Rev.}, vol.~55, pp. 999--1009, 2016.

\bibitem{Tamrakar:2017}
U.~Tamrakar, D.~Shrestha, M.~Maharjan, B.~P. Bhattarai, T.~M. Hansen, and
  R.~Tonkoski, ``Virtual inertia: Current trends and future directions,''
  \emph{Applied Sciences (Switzerland)}, vol.~7, no.~7, 2017.

\bibitem{NERC:2019}
{\relax North American Electric Reliability Corporation (NERC)}, \emph{2019
  Frequency Response Annual Analysis}, 2019.

\bibitem{7540970}
F.~M. {Mele}, A.~{Ortega}, R.~{Zarate-Minano}, and F.~{Milano}, ``Impact of
  variability, uncertainty and frequency regulation on power system frequency
  distribution,'' in \emph{Power Systems Computation Conference (PSCC)}, Jun.
  2016, pp. 1--8.

\bibitem{8626538}
P.~{Vorobev}, D.~M. {Greenwood}, J.~H. {Bell}, J.~W. {Bialek}, P.~C. {Taylor},
  and K.~{Turitsyn}, ``Deadbands, droop, and inertia impact on power system
  frequency distribution,'' \emph{{IEEE} Trans. Power Syst.}, vol.~34, no.~4,
  pp. 3098--3108, Jul. 2019.

\bibitem{GB:2016}
\emph{\text{National grid frequency data for 2016}},
  \relax{{https://www.nationalgrideso.com/\\
  balancing-services/frequency-response-services/historic-frequency-data}}.

\bibitem{6547228}
F.~{Milano} and R.~{Zárate-Miñano}, ``A systematic method to model power
  systems as stochastic differential algebraic equations,'' \emph{{IEEE} Trans.
  Power Syst.}, vol.~28, no.~4, pp. 4537--4544, Nov. 2013.

\bibitem{5298967}
R.~{Singh}, B.~C. {Pal}, and R.~A. {Jabr}, ``Statistical representation of
  distribution system loads using gaussian mixture model,'' \emph{{IEEE} Trans.
  Power Syst.}, vol.~25, no.~1, pp. 29--37, Feb. 2010.

\bibitem{574922}
A.~K. {Ghosh}, D.~L. {Lubkeman}, M.~J. {Downey}, and R.~H. {Jones},
  ``Distribution circuit state estimation using a probabilistic approach,''
  \emph{{IEEE} Trans. Power Syst.}, vol.~12, no.~1, pp. 45--51, 1997.

\bibitem{DELGADO2014267}
C.~Delgado and J.~Domínguez-Navarro, ``Point estimate method for probabilistic
  load flow of an unbalanced power distribution system with correlated wind and
  solar sources,'' \emph{Int. J. Electr. Power Energy Syst.}, vol.~61, pp.
  267--278, 2014.

\bibitem{Arnold1974}
L.~Arnold, \emph{Stochastic Differential Equations: Theory and
  Applications}.\hskip 1em plus 0.5em minus 0.4em\relax Wiley, 1974.

\bibitem{PhysRevE.54.2084}
\BIBentryALTinterwordspacing
D.~T. Gillespie, ``Exact numerical simulation of the ornstein-uhlenbeck process
  and its integral,'' \emph{Phys. Rev. E}, vol.~54, pp. 2084--2091, Aug. 1996.
  [Online]. Available: \url{https://link.aps.org/doi/10.1103/PhysRevE.54.2084}
\BIBentrySTDinterwordspacing

\bibitem{7731152}
H.~{Zhao}, Q.~{Wu}, S.~{Huang}, H.~{Zhang}, Y.~{Liu}, and Y.~{Xue},
  ``Hierarchical control of thermostatically controlled loads for primary
  frequency support,'' \emph{{IEEE} Trans. Smart Grid}, vol.~9, no.~4, pp.
  2986--2998, Jul. 2018.

\bibitem{4275780}
Z.~{Xu}, J.~{Ostergaard}, M.~{Togeby}, and C.~{Marcus-Moller}, ``Design and
  modelling of thermostatically controlled loads as frequency controlled
  reserve,'' in \emph{Procs.~of the IEEE PES General Meeting}, June 2007, pp.
  1--6.

\bibitem{6832599}
H.~{Hao}, B.~M. {Sanandaji}, K.~{Poolla}, and T.~L. {Vincent}, ``Aggregate
  flexibility of thermostatically controlled loads,'' \emph{{IEEE} Trans. Power
  Syst.}, vol.~30, no.~1, pp. 189--198, Jan. 2015.

\bibitem{8586247}
F.~Baccino, F.~Conte, S.~Massucco, F.~Silvestro, and S.~Grillo, ``{Frequency
  Regulation by Management of Building Cooling Systems Through Model Predictive
  Control},'' in \emph{2014 Power Systems Computation Conference (PSCC)}, 2014,
  pp. 1--7.

\bibitem{7579133}
V.~{Trovato}, I.~M. {Sanz}, B.~{Chaudhuri}, and G.~{Strbac}, ``Advanced control
  of thermostatic loads for rapid frequency response in great britain,''
  \emph{{IEEE} Trans. Power Syst.}, vol.~32, no.~3, pp. 2106--2117, May 2017.

\bibitem{jibji2019frequency}
F.~Jibji-Bukar and O.~Anaya-Lara, ``Frequency support from photovoltaic power
  plants using offline maximum power point tracking and variable droop
  control,'' \emph{IET Renewable Power Generation}, 2019.

\bibitem{fang2018improved}
J.~Fang, P.~Lin, H.~Li, Y.~Yang, and Y.~Tang, ``An improved virtual inertia
  control for three-phase voltage source converters connected to a weak grid,''
  \emph{{IEEE} Trans. Power Electron.}, vol.~34, pp. 8660--8670, Sep. 2019.

\bibitem{kim2019supercapacitor}
J.~Kim, V.~Gevorgian, Y.~Luo, M.~Mohanpurkar, V.~Koritarov, R.~Hovsapian, and
  E.~Muljadi, ``Supercapacitor to provide ancillary services with control
  coordination,'' \emph{{IEEE} Trans. Ind. Appl.}, vol.~55, no.~5, pp.
  5119--5127, 2019.

\bibitem{8052254}
F.~Bizzarri and A.~Brambilla, ``{PAN} and {MPanSuite}: Simulation vehicles
  towards the analysis and design of heterogeneous mixed electrical systems,''
  in \emph{IEEE International Conference of New Generation of Circuits and
  Systems (NGCAS)}, Sep. 2017, pp. 1--4.

\bibitem{Milano:2010}
F.~Milano, \emph{Power System Modelling and Scripting}.\hskip 1em plus 0.5em
  minus 0.4em\relax London: Springer, 2010.

\bibitem{Vancouver}
------, ``A {Python}-based software tool for power system analysis,'' in
  \emph{{Procs.~of the IEEE PES General Meeting}}, Vancouver, BC, Jul. 2013.

\bibitem{8727917}
M.~{Liu}, F.~{Bizzarri}, A.~M. {Brambilla}, and F.~{Milano}, ``On the impact of
  the dead-band of power system stabilizers and frequency regulation on power
  system stability,'' \emph{{IEEE} Trans. Power Syst.}, vol.~34, no.~5, pp.
  3977--3979, Sep. 2019.

\bibitem{Fate}
Y.~Zhang, W.~Yao, S.~You, W.~Yu, L.~Wu, Y.~Cui, and Y.~Liu, ``Impacts of power
  grid frequency deviation on time error of synchronous electric clock and
  worldwide power system practices on time error correction,'' \emph{Energies},
  vol.~10, pp. 1--15, 08 2017.

\bibitem{6855377}
P.~M. {Ashton}, C.~S. {Saunders}, G.~A. {Taylor}, A.~M. {Carter}, and M.~E.
  {Bradley}, ``Inertia estimation of the gb power system using synchrophasor
  measurements,'' \emph{{IEEE} Trans. Power Syst.}, vol.~30, no.~2, pp.
  701--709, Mar. 2015.

\bibitem{8469998}
R.~{Hollinger}, A.~M. {Cortes}, and T.~{Erge}, ``Fast frequency response with
  bess: A comparative analysis of germany, great britain and sweden,'' in
  \emph{15th International Conference on the European Energy Market (EEM)},
  Jun. 2018, pp. 1--6.

\bibitem{CONTE2017291}
F.~Conte, S.~Massucco, F.~Silvestro, E.~Ciapessoni, and D.~Cirio, ``Stochastic
  modelling of aggregated thermal loads for impact analysis of demand side
  frequency regulation in the case of sardinia in 2020,'' \emph{Int. J. Electr.
  Power Energy Syst.}, vol.~93, pp. 291--307, 2017.

\bibitem{TANG2019197}
Z.~X. Tang, Y.~S. Lim, S.~Morris, J.~L. Yi, P.~F. Lyons, and P.~C. Taylor, ``A
  comprehensive work package for energy storage systems as a means of frequency
  regulation with increased penetration of photovoltaic systems,'' \emph{Int.
  J. Electr. Power Energy Syst.}, vol. 110, pp. 197--207, 2019.

\bibitem{MAGDY2019351}
G.~Magdy, G.~Shabib, A.~A. Elbaset, and Y.~Mitani, ``Renewable power systems
  dynamic security using a new coordination of frequency control strategy based
  on virtual synchronous generator and digital frequency protection,''
  \emph{Int. J. Electr. Power Energy Syst.}, vol. 109, pp. 351--368, 2019.

\end{thebibliography}
